\documentclass[reprint,aps,prl]{revtex4-1}
\usepackage{graphicx}
\usepackage{pdfpages}
\usepackage{comment}
\usepackage{bm}
\usepackage{amsmath,amssymb}
\usepackage[utf8]{inputenc}
\usepackage{amsmath}
\usepackage{amssymb}
\usepackage{color}
\usepackage{subfigure}
\usepackage{ulem}
\usepackage{amsmath}
\usepackage{braket}
\usepackage{hyperref}

\makeatletter
\AtBeginDocument{\let\LS@rot\@undefined}
\makeatother

\begin{document}
\title{Optimal Configurations and ``Pauli Crystals" of Quantum Clusters} 
\author{Tin-Lun Ho and Saad Khalid}
\affiliation{Department of Physics, The Ohio State University, Columbus, OH 43210, USA} 
\date{\today}
\begin{abstract}
Broken rotational and translational symmetries are the hallmarks of solid state materials. In contrast, quantum liquids and gases do not exhibit such properties. However, if we regard the logarithm of the absolute square of a quantum liquid as an energy ${\cal E}= -{\rm ln}|\Psi|^2$, a geometric pattern naturally occurs at the minimum, i.e. the optimal configuration. Such geometric patterns have recently been studied for non-interacting fermions, and have been named ``Pauli crystals". However, such patterns exist in all interacting gases (Bose or Fermi), independent of statistics. Here, we present an algorithm to determine the optimal configurations of quantum clusters solely from the images of their densities and without theoretical inputs. We establish its validity by recovering a number of exact results, showing that it can identify the changes in the cluster's ground state which corresponds to phase transitions in bulk systems. 
\end{abstract}
\maketitle

According to the Copenhagen interpretation of quantum mechanics, the absolute square of the wavefunction $|\Psi[{\bf r}]|^2$  represents the probability of realizing the particle configuration $[{\bf r}]$ of a many-body system. In each single shot experiment, the  configuration realized is a set of points. The probability distribution $P([{\bf r}]) = |\Psi[{\bf r}]|^2$
only emerges after a large number of images are superposed together. This is best illustrated by the famous double slit experiment. 
In condensed matter, however, quantum mechanical averages are often observed directly in experiments. This is because most condensed matter measurements involve counting a large number of electrons, spins, or photons over some time interval, which amounts to accumulating many single shot measurements.
Quantum gases form a bridge between the condensed matter regime and the atomic regime. The number of particles in quantum gas experiments ranges from  $10^{1}$ to $10^6$. Interestingly, many important quantum phenomena such as antiferromangetism and quantum Hall states can also emerge in clusters with a few to a few tens of particles. With advances in cold atom experiments allowing one to to image the positions of individual atoms in a variety of settings\cite{Greiner1, Bloch1, Martin, Ott, Greiner2, Bloch2,Jochim0}, we can now access the probability distribution $P[{\bf r}]$ with these high resolution images.  

Hidden in  these images is the optimal configuration $[{\bf r}^{o}]$ that maximizes $P[{\bf r}]$. Not only is this the most probable region in configurations space, but it also presents a new way to detect quantum phase transitions, as it undergoes significant changes with the changes in the ground state. The optimal configurations (OCs) of the quantum clusters of free fermions in 2D harmonic traps have been studied\cite{Pol1, Pol2}. It is found that for small clusters up to around 7 particles, their OCs exhibit discrete $p$-fold symmetry with $p\leq N$. 
However, due to the continuous rotational symmetry of these systems, these OCs do not readily reveal themselves in the images of the clusters, since  all rotations of the same OC are equally probable.  To reveal the OC, the authors in reference\cite{Pol1, Pol2} have proposed the following ``alignment" scheme. 
One first finds the OC of a cluster using Monte Carlo method, (a theoretical input), and then  rotates each experimental image by an angle so as to minimize the ``angular distance" between the  rotated image and the numerically found OC.  It is claimed that the true OC will emerge after the data is so processed\cite{Pol1, Pol2}. 
This method has been  implemented recently by Salim Jochim's group\cite{Salim}, and the predicted OCs do appear.  However, there are two problems. Firstly, this method relies on a theoretical input. If such input is unavailable, such as for most interacting systems, then one will not know how to start.
Secondly, this alignment algorithm\cite{Pol1, Pol2} was seriously challenged in Ref.\cite{challenge}. It is shown that even a sample of random configurations after processed through this algorithm will produce a pattern matching the reference pattern, whatever it is. The authors of ref.\cite{challenge} suggest that the results of the alignment algorithm are artificial. Responding to this challenge, the authors in ref.\cite{Salim} have provided more evidence in their Supplementary Material to support  the intrinsic nature of their findings. This issue aside, the 
alignment algorithm\cite{Pol1, Pol2} still relies on knowing the answer in the first place, which are not forthcoming in general. It is therefore desirable to have a method to determine the intrinsic geometric structure of a quantum cluster, based solely on the experimental data and without the inputs of specific theories. We shall present such an algorithm below, and demonstrate its validity by recovering the exact results of a number of benchmark systems. 

Due to the Pauli exclusion principle, the images of a cluster of $N$ spinless fermions  ($N$-cluster for short) is a set of $N$ discrete points. Since the OCs of small clusters are regular polygons\cite{Pol1, Pol2}, they are given the name ``Pauli crystals". This terminology is misleading because these patterns also exist in interacting Bose clusters, independent of statistics.  Nor are these patterns crystals in any conventional sense. We mention this term in the title only for historical references. We shall use ``optimal configuration" (OC) from now on for accuracy. 

\begin{figure}
\centering
\includegraphics[width=3.2in, height=2.5in]{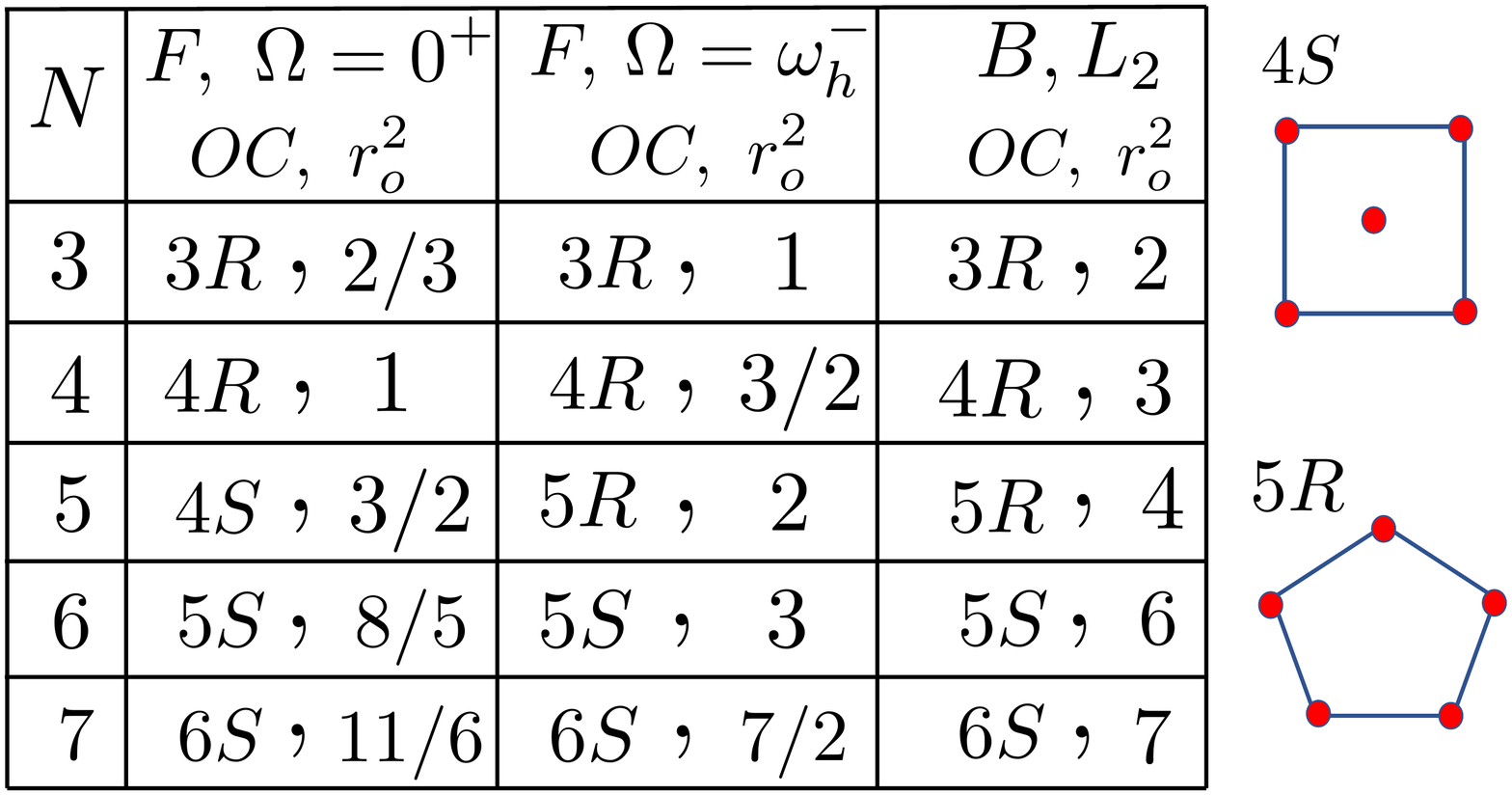}
\caption{The first two columns are the OCs of spinless fermions in the slow rotation $(\Omega=0^{+})$ and and fast rotation $(\Omega= \omega_{h}^{-})$ limit. The last column are bosonic Laughlin state $L_{2}$ with filling factor 1/2. The 5-ring and 4-star configurations for a 5-cluster are shown on the right. All lengths are in units of harmonic oscillator length $d=\sqrt{\hbar/(M\omega_{h})}$. See also Supplementary Materials (SM).  }
\label{fig1}
\end{figure}


{\bf I. The OCs and the principle curvatures:}   To connect to current experiments\cite{Salim}, we shall consider a quantum gas in a 2D rotating harmonic trap with trap frequency $\omega_{h}$  and rotating frequency $\Omega$. The hamiltonian is 
$H= \sum_{i]1}^{N}h_{i} + V$, where $h_{i} = \big( \frac{{\bf p}_{i}^{2}}{2M} + \frac{1}{2}M\omega^2_{h} {\bf r}_{i}^2 - \Omega \hat{\bf z}\cdot {\bf r}_{i}\times {\bf p}_{i} \big) $, and  $V=g\sum_{i>j} \delta({\bf r}_{i} - {\bf r}_{j})$ is the interaction between particles (spinful fermions or interacting bosons). 
We shall denote the particle configurations $({\bf r}_{1}, {\bf r}_{2}, ..., {\bf r}_{N})$ by an $N$-dimensional complex vector ${\bf z} = (z_1, z_2, .., z_N)$.  We shall from now on expressed all lengths in units of the harmonic oscillator length $d= \sqrt{\hbar/(M\omega_{h})}$. 
Since the center of mass (CM) coordinate ${\cal Q} = \sum_{i=1}^{N} z_{i}/N$
can be separated out in $H$, the ground state then satisfies $\Psi({\bf z}) = e^{-N|{\cal Q}|^2/2}\Psi ({\bf w})$, where $w_{i} = z_{i} - {\cal Q}$ is the particle coordinates in the CM frame. For later discussions, we define a dimensionless ``energy" ${\cal E}({\bf z})$ as
$P({\bf z})\equiv |\Psi({\bf z})|^2 = e^{- {\cal E({\bf z})}}$. It satisfies
${\cal E}({\bf z}) =  {\cal E}({\bf w}) + N|{\cal Q}|^2$. 
The OC  of an $N$-cluster is the is the configuration ${\bf z}^{o}$ that
minimizes the ``energy" ${\cal E}({\bf z})$,  $\left( \partial {\cal E}/\partial z_{i}\right)_{{\bf z}^{o}} = 0$.
Near the minimum ${\bf z}^{o}$, the ``energy"  increases as $\delta {\cal E} =  \frac{1}{2}{\cal Z}^{\dagger}_{i} {\cal W}_{ij} {\cal Z}_{j}$, 
where 
\begin{equation}
{\cal Z}_{i} = \left( \begin{array}{c} \delta z_{i} \\ \delta z_{i}^{\ast}\end{array} \right), \,\,\,\, 
 {\cal W}_{ij} =  
  \left( \begin{array}{cc} 
    M_{ij} & Q_{ij} \\ Q^{\ast}_{ij} & M^{\ast}_{ij}   \end{array} \right)
\label{deltaE} \end{equation}
 where $\delta z_{i} = z_{i} - z^{(o)}_{i}$, 
 $M_{ij} =\partial_{z_{i}^{\ast}}\partial_{z_{j}} {\cal E}({\bf z}^{o})$, and $Q_{ij} =\partial_{z_{i}^{\ast}}\partial_{z_{j}^{\ast}} {\cal E}({\bf z}^{o})$. We shall denote 
 the eigenvalues and eigenvectors of ${\cal W}$ as $\lambda_{n}$ and $({\bf v}^{(n)}, {\bf v}^{(n)\ast})$ with normalization $|{\bf v}_{n}|=1$. For displacements $\delta {\bf z} = \rho  {\bf v}^{(n)}$ away from the OC, the probability decreases as 
 \begin{equation}
    P({\bf z}^{o}+ \rho {\bf v}^{(n)})/P({\bf z}^{o}) = e^{- \lambda_{n}
    \rho^2}.
\label{cur} \end{equation}
 ${\bf v}^{(n)}$ and $\lambda_{n}$ will be referred to as the principal vectors and principal curvatures of the OC. Note that Eq.(\ref{cur}) also applies to the CM frame. 
 
\begin{figure}
\centering
\includegraphics[width=3.2in, height=2.5in]{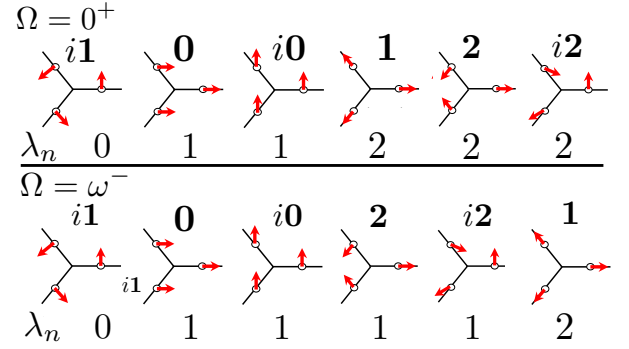}
\caption{The principal vectors ${\bf v}^{(n)}$ and the principle curvatures $\lambda_{n}$ of a $N=3$ cluster of spinless fermion in slow rotation $\Omega=0^{+}$ and in quantum Hall regime $\Omega=\omega^{-}_{h}$. In both cases, the OC is a 3-ring. The numerals are the values of $\lambda_{n}$ for the mode depicted. 
The bold letters {\bf 1} to {\bf 3} denotes ${\bf v}^{(1)}$ to ${\bf v}^{(3)}$,  ${\bf v}^{(n)} = (1, \omega^{n}, \omega^{2n} )/\sqrt{3}$, 
$n=0,1,2$, and $\omega = e^{i 2\pi/3}$. The symbol  
 i{\bf 2} means the vector $i {\bf v}^{(2)}$.  See also (SM).}
\label{fig2}
\end{figure}

 {\bf II. The OCs of slow and fast rotating  clusters:} The ground states of interacting quantum clusters are  difficult to obtain, as exact numerical diagonalization can only handle very few particles. On the other hand, the ground states of spinless fermions as well as some quantum Hall ground states of bosons and fermions are known. 
 Consequently, one can determine their OCs and their principle curvatures by minimizing the ``energies" ${\cal E}$ and diagonalizing the matrix ${\cal W}$ in Eq.(\ref{deltaE}). 
 The results up to 7 particles are shown in Fig.\ref{fig1}. The OCs of these clusters are either $N$-rings (a regular $N$-polygon), or $N-1$-stars (an $(N-1)$-ring plus a point at the center). The optimal radius of the polygon is denoted as $r_{o}$. In Fig.\ref{fig2}, we show the principle curvatures $\lambda_{n}$ and the principle vectors of a 3-cluster in the  slow rotation ($\Omega=0^{+}$) and quantum Hall $(\Omega= \omega^{-}_{h})$ regimes. The  derivations of these results are given in the Supplementary Material (SM), where we also point out that the principle vectors of an $N$-ring are determined entirely by symmetry (independent of theory). 
 These exact results will serve as benchmarks for any proposal to deduce the OCs from the experimental images.
 It is also clear from Figure 1 and 2  that the OCs and their principle curvatures change significantly as the  ground state changes. They can therefore be used to detect quantum phase transitions in the bulk.  

{\bf III. To determine the OC solely from density images:} Our algorithm is designed for interacting clusters with discrete rotational symmetry. 
Without loss of generality, we only consider configurations in the CM frame, which can be obtained from the observed images by subtracting off their CM coordinates. 
We  denote a sample of  $M$  images of an $N$-cluster in the CM frame  as ${\cal S}= ({\bf w}^{(1)},
{\bf w}^{(2)}, ..., {\bf w}^{(M)})$, $\sum_{i=1}^{N}w_{i}^{(p)} = 0$, $p=1, .. , M$. 
Before presenting our scheme, let us first explain the  difficulties of measuring the probability $P({\bf w})$ according to  its definition. Let ${\cal N}_{\epsilon}({\bf w}; {\cal S})$ (referred to as  ``correlation number") be the $N$-point correlation of the ground state averaged over a  neighborhood of ${\rm d }{\bf w}$ around ${\bf w}$, 
\begin{equation}
{\cal N}_{\epsilon}({\bf w}; {\cal S}) = \sum_{\alpha=1}^{M} \prod_{i=1}^{N}\Theta_{\epsilon}( {\bf z}^{(\alpha)} - {\bf w}), 
\label{N}\end{equation}
where $\Theta_{\epsilon}( {\bf z}^{(\alpha)} - {\bf w}) \equiv
\prod_{i=1}^{N} \Theta_{\epsilon} (z^{(\alpha)}_{i}- w_{i})$, and $\Theta_{\epsilon} (z)$ is a disc of radius $\epsilon$ surrounding $z$; $\Theta_{\epsilon} (z)=1$ if $|z|<\epsilon$, and 0 otherwise.  See Fig.3(a). If ${\cal S}$ is a  large set, then we have 
\begin{equation}
P({\bf w}) = \underset{\epsilon \rightarrow 0}{ {\rm Lim}}\underset{M\gg 1}{ {\rm Lim}}{\cal N}_{\epsilon}({\bf w}; {\cal S})/( M (\pi \epsilon^2)^N)
\label{P} \end{equation}
where $(\pi \epsilon^2)^N$ is the volume of ${\rm d}{\bf w}$, and $\int {\rm d}{\bf w}P({\bf w})=1$.  If one could measure ${\cal N}_{\epsilon}({\bf w}; {\cal S})$ for all configurations ${\bf w}$, then the OC is simply the configuration with the largest correlation number ${\cal N}_{\epsilon}({\bf w}; {\cal S})$. 
However, this is impractical. Firstly, the number ${\cal N}_{\epsilon}({\bf w}; {\cal S})$ is very small. Measuring it according to Eq.(\ref{P})  will have poor signal to noise ratio.  
Secondly, the number of configurations is infinite. It is impossible to measure the probabilities of all configurations. 

\begin{figure}
\centering
\includegraphics[width=3.2in, height=2.5in]{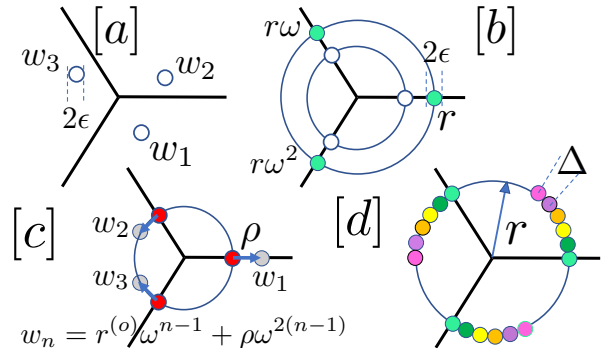}
\caption{(a) Measurement of the correlation number ${\cal N}_{\epsilon}({\bf w}, {\cal S})$  for an arbitrary configuration ${\bf w}=(w_1, w_2, w_3)$ of a 3-cluster: According to Eq.(\ref{N}), (\ref{P}),  ${\cal N}_{\epsilon}({\bf w}, {\cal S})$ is the experimental images that fall into the three circles of radius $\epsilon$ around $w_{i}$. This  number is largest for the OC. 
(b) Once the symmetry of the OC is determined using the method in ${\bf (IIIA)}$, (say, a 3-ring for a 3-cluster), 
then the search of the OC reduces to searching within the 3-rings family $r{\bf w}_{R}=r(1,\omega, \omega^2), \omega=e^{2\pi i/3}$. 
An efficient way to evaluate the probability $P(r)=P(r{\bf w}_{R})$ is described in Section {\bf (IIIB)}. The OC is the maximum of $P(r)$, 
with optimal radius $r_{o}$. 
(c) Once the OC (the set of red circles) is determined, one can move away from the OC along the directions of one of the principle vectors. What is depicted is the deviation ${\bf w}= r_{o}{\bf w}_{R}+ \rho {\bf v}^{(2)}$, where ${\bf v}^{(2)}$ is the same ${\bf v}^{(2)}$ mode in Fig.\ref{fig2}. 
Calculating the probability of this deviation with the method in Section {\bf (IIIB)}, we can deduce the principle curvature $\lambda_{2}$ using  Eq.(\ref{cur}). 
(d) Derivation of Eq.(\ref{PPR}): To increase the accuracy of the measurement of ${\cal N}_{\epsilon}({\bf w}, {\cal S})$, we imagine a large number of measurements performed on a sequence of rotated rings (indicated in different colors). The number of such measurements is $2\pi r/\Delta$, and the sum of these measurement is 
$(2\pi r/\Delta ){\cal N}_{\epsilon}({\bf w}, {\cal S})$, which is also the correlation number of the x-aligned sample $\widetilde{\cal S}_{x}$ as discussed in Section  {\bf (IIIB)}.  }
\label{fig3}
\end{figure}

\begin{figure*}
\centering
\includegraphics[width=.7\textwidth]{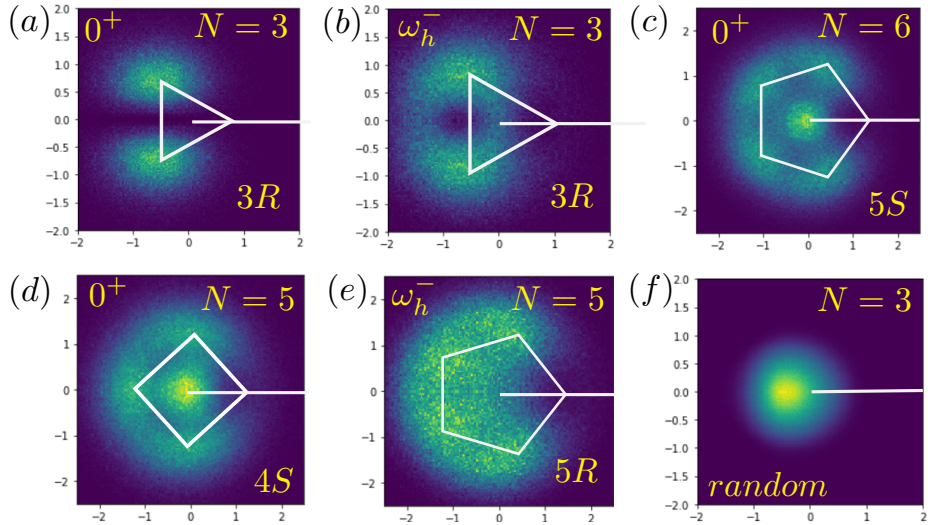}
\caption{(a)-(e) are the superposition of the configurations in the $x$-aligned sample ${\cal S}_{x}$ of an $N$-cluster ($N=3,6$) of spinless fermions in the slow rotation $(\Omega=0^{+})$ and fast rotation limit $(\Omega=\omega^{-}_{h})$. The sample ${\cal S}_{x}$ is generated from an original sample ${\cal S}$ consisting of $M$=20K configurations. For OCs that are $N$-rings, there are $N-1$ blobs on the symmetry axes of the ring (excluding the x-axis). There are $NM$ points on the positive x-axis. They are not shown in the figure. For an $N$-star, a blob also shows up at the center. The symmetry of the OCs deduced from these superposition agree with the results in Table 1.  (f) is the superposition of the $x$-aligned sample ${\cal S}_{x}$ of  random configurations of a 3-cluster. It shows no particular symmetry. The off center bright spot is due to the fact that the configurations are in the CM frame. }
 \label{fig4}
\end{figure*}

\begin{figure}
\centering
\includegraphics[width=3.2in, height=2.5in]{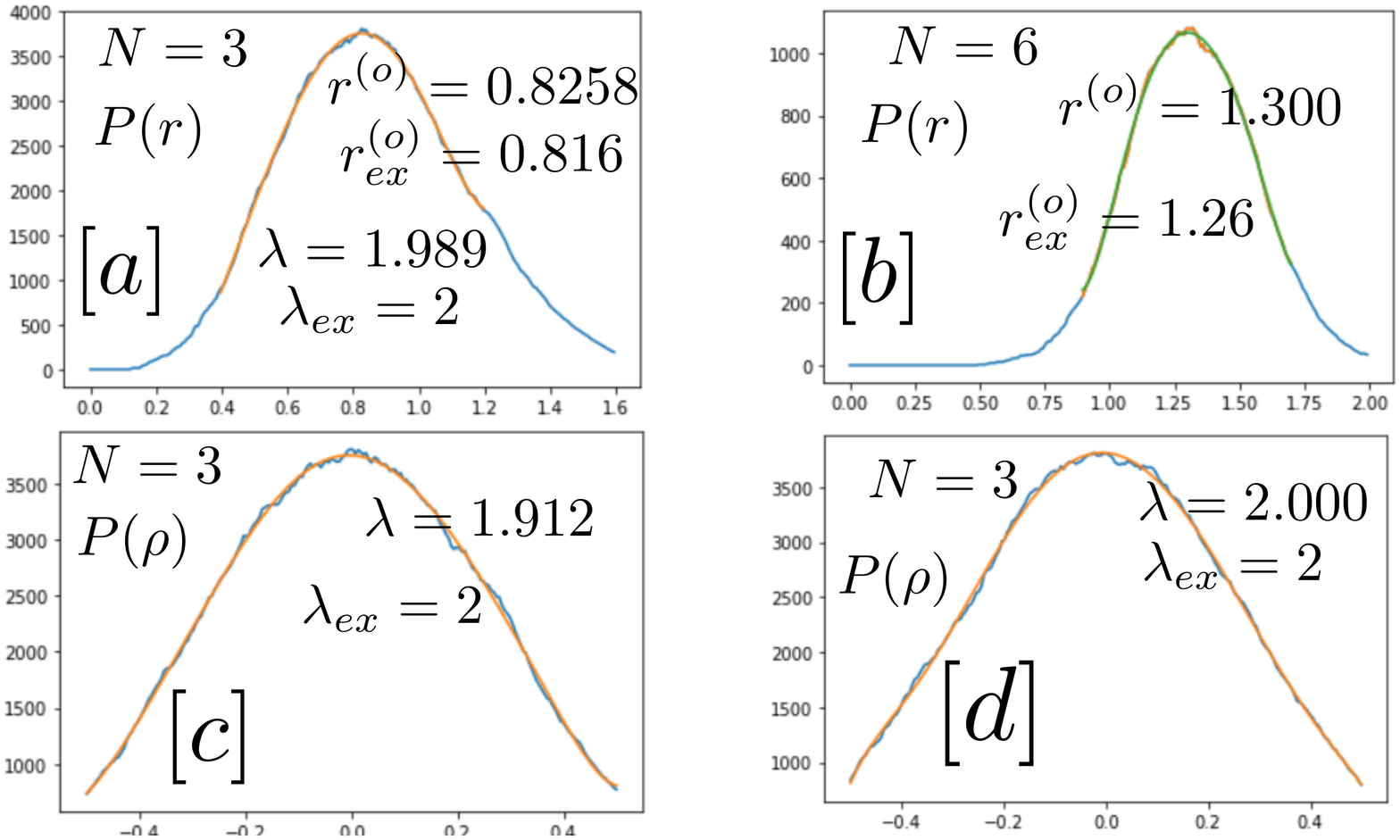}
\caption{  (a) and (b) are the probability $P(r)$ of a 3-cluster and a 6-cluster of spinless fermions in low rotation limit $(\Omega=0^{+})$, calculated from  from Eq.(\ref{PPR}) using an original sample  ${\cal S}$ with $M=500K$. The radius of the collection disc is $\epsilon=0.12$, a small size yet big enough to produce a smooth curve $P(r)$. The maximum radii $r_{o}$ are determined by a power law fit around the maximum. They differ from the exact value $r_{ex}$ by a few percent.  The stretching mode along the symmetry direction as well as other two modes of the 3-clusters have also been calculated using Eq.(\ref{PPR}) and Eq.(\ref{cur}) as described in the text. The principle curvatures $\lambda$ of these modes are given in Fig.(a)-(d). They agree with the exact values $\lambda_{ex}$ in Fig.\ref{fig2} by a few percent. In (c) and (d), the deviation from the OC is given by $\delta{\bf w}= {\bf w}- r_{o}{\bf w}_{R} $, where $\delta{\bf w}=\rho{\bf v}^{(2)}$ and $i \rho {\bf v}^{(2)}$ are the vectors shown in Fig.\ref{fig2}. For a sample with $M$=50K configurations, the error increases to about $10\%$, but still smaller than the difference of two optimal radii of the $N=3$-fermion cluster in Fig.\ref{fig1}.}
\label{fig5}
\end{figure}

Here, we show that both problems can be solved by constructing of a new sample ${\cal S}_{x}$ as follows. For each of the $M$ configurations (say, ${\bf w}=(w_1, w_2, ..., w_{N})$) in ${\cal S}$, we perform a rotation $\eta_{i}$ so that the  point $w_i$ will be rotated onto the positive $x$-axis in the rotated configuration $\eta_{i}{\bf w}^{}$; (i.e. $\eta_{i}\equiv e^{-i \theta_{i}}$, and $\theta_{i}= {\rm arg} (w_{i})$).   Applying this procedure to other $N-1$ points in ${\bf w}$, we generate altogether $N$ such rotated configurations, which we  call the ``$x$-aligned" configurations. The original sample ${\cal S}$ now becomes an ``$x$-aligned sample"  ${\cal S}_{x}$ consisting of $NM$ configurations.
What ${\cal S}_{x}$ does is to bring close together those rotational equivalent configurations in  ${\cal S}$ that are originally far apart (i.e. with little overlaps). 
As a result, the correlation number of the sample ${\cal S}_{x}$ is significantly larger than that of ${\cal S}$.  With the ``x-aligned" sample, the OC can be determined in two steps. 

\noindent {\bf (IIIA)} {\em To reveal the hidden symmetry}: We first superpose all $NM$ images of ${\cal S}_{x}$ together. If the OC is a $\ell$-ring, 
the superposed pattern will have $NM$ points on the $x$-axis (which is one of the symmetry axis) and $\ell-1$ blobs along the other $N-1$ symmetry axes. 
If the OC is a star, an additional blob will also show up at the center. 
We illustrate these phenomena in Fig.\ref{fig4} for a 3-cluster and a 6-cluster of spinless fermion in both slow rotation and in quantum Hall regime. For each of these clusters, we  generate a sample ${\cal S}$ with $M=20K$ configurations from their ground state wavefunctions. These configurations will be considered as the ``experimental" images of these clusters. Fig.\ref{fig4} is the superposition of the ``x-aligned" images ${\cal S}_{x}$ of these clusters. The underlying symmetries of their OCs are identified by the number of bulbs that appear. These symmetries agree with those in Fig.\ref{fig1}. Applying this procedure to a sample of random configurations shows no particular symmetry.

\noindent {\bf (IIIB)} {\em Finding the OC from a new formula for the probability $P(r)$:}
Once the symmetry of the OC is identified, (say, an $N$-ring), it remains to evaluate the probability $P(r)=P(r{\bf w}_{R})$
within the $N$-ring family 
\begin{equation}
  r{\bf w}_{R}= r(1,\zeta, \zeta^2, .., \zeta^{N-1})/\sqrt{N}, \,\,\,\, \zeta= e^{2\pi i/N}, 
  \label{ring}
\end{equation} 
and to determine the optimal radius $r_{o}$. (See Figure 2b). 
Our key result is that 
$P(r)$ is proportional to the correlation number ${\cal N}_{\epsilon}(r{\bf w}_{R}; {\cal S}_{x})$ of the x-aligned sample {\em  divided by $r$.} 
  \begin{equation}
      P(r)  \propto  \underset{\epsilon \rightarrow 0}{ {\rm Lim}} \frac{{\cal N}_{\epsilon}({\bf w}(r); {\cal S}_{x})}{r}, \,\,\,\,\, {\bf w}(r)= r{\bf w}_{R}. 
  \label{PPR}
  \end{equation}
The division by $r$ is crucial.  
The OC is $r_{o}{\bf w}_{R}$, where $r_{o}$ maximizes $P(r)$
The OC is the maximum $r_{o}$ of $P(r)$. 
Once the OC ${\bf w}_{o}$ is  determined, Eq.(\ref{PPR}) can also be used to find the  probabilities for configurations ${\bf w'}$ close to the OC by  replacing 
the configuration ${\bf w}(r)= r{\bf w}_{R}$ in Eq.(\ref{PPR}) by ${\bf w}'$. If the deviation ${\bf w'}- {\bf w}_{o}$ is chosen along  one of the principle direction, (see Figure 2c), we can then evaluate the  ratio in Eq.(\ref{cur}) and find the corresponding principle curvatures.  In the following, we shall
first demonstrate the validity of Eq.(\ref{PPR}),  and then return to its proof.

In Fig.5a and 5b, we show the curves $P(r)$ for the 3-cluster and 6-cluster of spinless fermions in the low rotation limit. (The symmetries of these clusters have been determined as  $3R$ and $5S$ in Figure 4.) Theses curves are calculated for a very large sample with $M$=500K configuration with a disc size $\epsilon=0.12$. (This disc size is chosen so that $\epsilon$ remains small but large enough to produce a smooth curve $P(r)$). 
The optimal radii $r_{o}$ determined from the peak of the distribution match the exact value $r_{ex}^{o}$ up to a few percent. For both $N=3$ and $N=6$ clusters, the principle curvature associated with the stretching mode along the symmetry axes can be determined using Eq.(\ref{PPR}) with ${\bf w}(r)$ replaced by 
$(r_{o}+\rho){\bf w}_{R}$. Again, the accuracy of this principle is about a few percent. See Figure 5a and 5b. Other principle values for the 3-cluster are shown in Figure 5c and 5d, again with similar accuracy. They are determined using the method shown in Figure 3c, and making use of Eq.(\ref{PPR}) and Eq.(\ref{cur}). 
 If a smaller sample with $M$=50K configurations is used, the errors for the optimal radii will increase to about 10 percent for the same circle size $\epsilon=0.12$. This error, however, is still less than the difference between in optimal radii of the 3-cluster in slow rotation and in the quantum Hall regime, (see Table 1).  The determination of the OC of the 3-clusters can therefore be used to detect the changes in the ground state. 
 
\noindent {\em Derivation of Eq.(\ref{PPR})}: We begin with Eq.(\ref{P}). To increase the signal to noise of the correlation number ${\cal N}({\bf w}; {\cal S} )$,  we make use of the rotational symmetry of $P({\bf w})$. Imagine that we perform the same measurement on a sequence of rotated rings $\zeta {\bf w}$ as shown in Figure 3d, where $\zeta$ is a small rotation (i.e. a small phase factor).  Rotational symmetry implies that  for  large samples $M\gg1$, 
\begin{eqnarray}
    P(r) = Q^{-1}\sum_{p=1}^{Q-1}{\cal N}_{\epsilon} ( \zeta^{-p} {\bf w}(r); {\cal S})/(M (\pi \epsilon^2)^{N})     \label{sum1}  \\
    =\sum_{p=0}^{Q-1}\sum_{\alpha=1}^{M} \Theta_{\epsilon}(\zeta^{p} {\bf z}^{(\alpha)} - {\bf w}(r))/(QM (\pi \epsilon^2)^{N}),   \label{sum2}
\end{eqnarray}
where $Q$ is the number of the rotated rings, and ${\bf w}(r)=r{\bf w}_{R}$ is a configuration on the $N$-ring family Eq.(\ref{ring}).   If the arc length between the different N-rings is $\Delta$, ( asmall number independent of $\epsilon$), then we have $Q= 2\pi r/\Delta $. See Figure 3d. Next, we note that the sum in Eq.(\ref{sum2}) can be rewritten as $\sum_{\beta=1}^{MQ} \Theta_{\epsilon}({\bf y}^{(\beta)} - {\bf w})$, where  $(\beta)\equiv (p, \alpha)$ and 
${\bf y}^{(\beta)} \equiv \zeta^{p} {\bf z}^{(\alpha)}$. This is 
precisely the correlation number of a sample $\widetilde{\cal S}$ with $QM$ configurations obtained by combining all the rotated configurations of ${\cal S}$. Hence we have 
\begin{equation}
     P(r) = \underset{\epsilon \rightarrow 0}{ {\rm Lim}}\underset{QM\gg 1}{ {\rm Lim}}\frac{ {\cal N}_{\epsilon}({\bf w}(r); \widetilde{\cal S})}{2 \pi r}
     \frac{\Delta }{M (\pi \epsilon^2)^{N} }. 
\end{equation}
Note that our $N$-ring $r{\bf w}_{R}$ contains the $x$-axis. Hence,   
in the limit of small $\epsilon$, only those configurations in $\widetilde{\cal S}$ that have a point on the $x$-axis will contribute to the sum.  The  number ${\cal N}({\bf w}(r); \tilde{\cal S})$ is therefore the same as that for the aligned sample ${\cal S}_{x}$, despite their different number of configurations.  
We then have Eq.(\ref{PPR}). 

{\em Concluding Remarks:} We have presented an algorithm to deduce the optimal configuration of a quantum cluster solely from its images, and have demonstrated its validity by recovering many exact results. The algorithm can be applied to interacting systems where exact solutions are unavailable. The images of a quantum cluster contain huge amount of information about its quantum coherence and entanglement.  
Our work shows that valuable information can be extracted from them by using a specially designed algorithm. There is much more information in these images, still waiting to be revealed. 

{\em Acknowledgments:} The work is supported by the MURI Grant FP054294-D.

\newpage
\onecolumngrid
\section*{Supplementary Material}

\begin{center}
{{\em Optimal Configurations and ``Pauli Crystals" of Quantum Clusters} by Tin-Lun Ho and Saad Khalid}  

\vspace{0.2in}

For the optimal configurations of small clusters -- the results are in Figures 1 and 2 :
\end{center}

\vspace{0.2in}

\noindent {\bf I.  Quantum clusters in rotating harmonic trap: }

The Hamiltonian is $H= \sum_{i]1}^{N}h_{i} + V$, where $h_{i} = \big( \frac{{\bf p}_{i}^{2}}{2M} + \frac{1}{2}M\omega^2_{h} {\bf r}_{i}^2 - \Omega \hat{\bf z}\cdot {\bf r}_{i}\times {\bf p}_{i} \big) $, and   $V=g\sum_{i>j} \delta({\bf r}_{i} - {\bf r}_{j})$ is the interaction between particles (spinful fermions or interacting bosons).  $\omega_{h}$ is the frequency of the harmonic trap, and $\Omega$ is the rotational frequency of the trap. 
The eigenfunctions and  eigen-energies of the single particle hamiltonian are 
\begin{equation} 
u_{n,\ell} ({\bf r}) = \sqrt{  \frac{n!}{(n+\ell)!  \pi} } 
z^{\ell} L_{n}^{(\ell)}(|z|^2) e^{-|z|^2/2},  
\,\,\, E_{n, \ell}=   \hbar(2n\omega + (\omega-\Omega)\ell), \,\,\, n+\ell \geq 0, 
\end{equation}
where ${\bf r}= (x,y)$, $z= r e^{i \varphi}/d = (x+iy)/d$,  $d=\sqrt{\hbar/(M\omega_{h})}$ is the harmonic oscillator length, 
$n\geq 0$ is a non-negative integer, and $\ell$ is an integer (both positive and negative) with the constraint $n+\ell\geq 0$, and $L_{n}^{(\ell)}(z)$ is the generalize Laguerre polynomial.  \\

The spectra of the single particle energy level in the slow rotation ($\Omega=0^{+}$) and in the fast rotation 
($\Omega=\omega_{h}^{-}$) cases are shown in Figure(SM1).  We shall label the single particle states with increasing energy as 1,2, 3, ... etc; and their eigenstates ($z^{\ell} L_{n}^{(\ell)}(|z|^2)$) as  $ f_1, f_2, f_3, ... $ etc.  The leading terms of these polynomials $f_{i}$ are also shown in Figure (SM1). The wavefunctions  $ f_1, f_2, f_3, ... $ in the fast rotating regime are the analytic functions $1, z, z^2, ..$. 
The ground state of N non-interacting fermions is a Slater determinant of the lowest $N$ states of $H$, of the form
\begin{equation}
   \Psi({\bf z}) = D({\bf z}) e^{- \sum_{i=1}^{N}|z_{i}|^2/2},
\end{equation}
where $ D({\bf z})$ is the determinant 
\begin{equation}
 D({\bf z}) = || f_1, f_2, .. f_N || \equiv \sum_{{\cal P}} (-1)^{{\cal P}} f_{1}(z_{{\cal P}1}) f_{2}(z_{{\cal P}2}).. f_{N}(z_{{\cal P}N}). 
\end{equation}
It is easy to show that in the ground state, the eigenstates $z^{\ell}L^{(\ell)}_{n}(|z|^2)$ in $D$ can be replaced by their leading term  $z^{\ell}|z|^{2n}$ with $\ell\leq n$. For example, the  determinants of a four particle and a five particle cluster in the low rotational limit ($\Omega= 0^{+}$) are $D_{4}({\bf z})= ||1, z, z^{\ast}, z^2||$
$D_{5}({\bf z})= ||1, z, z^{\ast}, z^2, z^{\ast}z||$, i.e.
\begin{equation}
    D_{4}(z_1, .., z_4) = \left| \begin{array}{cccc} 1 & z_1& z^{\ast}_{1} & z^{2}_{1} \\
    1 & z_2& z^{\ast}_{2} & z^{2}_{2}  \\
    1 & z_3& z^{\ast}_{3} & z^{2}_{3} \\
    1 & z_4& z^{\ast}_{4} & z^{2}_{4} 
    \end{array}
    \right|.
    \label{D4}
    \end{equation}

\begin{equation}
    D_{5}(z_1, .., z_5) = \left| \begin{array}{ccccc} 1 & z_1& z^{\ast}_{1} & z^{2}_{1} & z^{\ast}_{1} z_{1} \\
    1 & z_2& z^{\ast}_{2} & z^{2}_{2} & z^{\ast}_{2} z_{2} \\
    1 & z_3& z^{\ast}_{3} & z^{2}_{3} & z^{\ast}_{1} z_{3} \\
    1 & z_4& z^{\ast}_{4} & z^{2}_{4} & z^{\ast}_{4} z_{4}\\
    1 & z_5& z^{\ast}_{5} & z^{2}_{5} & z^{\ast}_{5} z_{5}
    \end{array}
    \right|.
    \label{D5}
\end{equation}

For fast rotations, ($\Omega= \omega_{h}^{-} = \omega_{h} - 0^{+} $), all the particles reside in the lower Landau level. For spinless fermions, the ground state is the Fermi sea with the lowest filled $N$ states $z^{\ell}, \ell=0, 1, .. , N-1$, which is Vendermont determinant $V({\bf z})= || 1, z, z^2, .., z^{N-1}||$.  For bosons, repulsive interaction can lead to  fractional quantum Hall states such as  the Laughlin state and the Pfaffian state, with wavefunctions
\begin{eqnarray}
 {\rm Laughlin} \,\, L_{m} \,\,\,\, &  D({\bf z}) = V({\bf z})^{m}  \,\,\,\,  \label{Laughlin} \\
 {\rm Pfaffian} \,\, Pf_{m} \,\,\,\, &  D({\bf z}) = V({\bf z})^{m} {\rm Pf}[(z_i-z_j)^{-1}
   \,\,\,\,   \label{Pfaffian}
\end{eqnarray}
where $V({\bf z})= || 1, z, z^2, .., z^{N-1}||$ is the 
Vendermont determinant, and $m$ an integer consistent with statistics. 
The optimal configuration is the minimum of the ``energy"  
\begin{equation}
{\cal E}({\bf z})= - {\rm in}|D({\bf z})|^2 + \sum_{i=1}^{N} |z_{i}|^2. 
\label{EE} \end{equation}
Although the minimum of Eq.(\ref{EE}) can be found numerically, the minimum of small clusters with discrete rotational symmetries can be obtained  analytically. For example, the configurations of $N$-ring and $(N-1)$-star families are given by 
\begin{eqnarray}
{\bf z} = r (1, \zeta, \zeta^2,  ... , \zeta^{N-1}), & \,\,\,\,\, \zeta = e^{2\pi i/N}    , & \,\,\,\, N-{\rm ring},    \label{Ring}\\
{\bf z} = r (0,  1, \eta, \eta^2,  ... , \eta^{N-2}),& \,\,\,\,\, \eta = e^{2\pi i/(N-1)}, & \,\,\,\, (N-1)-{\rm star}.     \label{Star}
\end{eqnarray}
Take the 4-cluster as an example. With the ground state in the slow rotation limit given in Eq.(\ref{D5}), substituting Eq.(\ref{Ring}) into the energy Eq.(\ref{EE}) gives  
\begin{eqnarray}
{\cal E}_{\rm 4-ring} (r) = - {\rm ln} r^8 - {\rm ln}|\Delta_{\rm 4-ring}(\zeta)|^2 + 4r^2, \,\,\,\,        \\
\Delta_{\rm 4-ring}(\zeta)=      \left| \begin{array}{cccc} 1 & 1& 1 & 1 \\
    1 & \zeta & \zeta^{\ast} & \zeta^{2}  \\
    1 & \zeta^2 & \zeta^{\ast 2} & \zeta^{4} \\
    1 & \zeta^3 & \zeta^{\ast 3} & \zeta^{6} 
    \end{array}
    \right|, \,\,\,\,\,  \zeta= e^{2\pi i/4}
    \label{E4ring}
    \end{eqnarray}
 Eq.(\ref{E4ring}) implies that the minimum is at  $r_{o}^2=1$.  \\  

For a 5-cluster, the wavefunction Eq.(\ref{D5}) vanishes on a 5-ring, since the column ``$zz^{\ast}$" reduces to $r^2$, which is proportional to the first column ``1".  {\em Hence, there are no $N$-ring optimal configurations for $N>4$, as shown in Figure 1}. In the family of 4-star (Eq.{Star}), the determinant Eq.(\ref{D5}) becomes
\begin{equation}
    D_{4-star}(0, r, r \zeta, r\zeta^2, r\zeta^3) = \left| \begin{array}{ccccc} 1 & 0 & 0 & 0 & 0 \\
    1 & r  &  r   &  r^2 \zeta^2 & r^2 \\
    1 & r \zeta & r \zeta^{\ast}  &  r^2 \zeta^2 & r^2 \\
    1 & r \zeta^2 & r \zeta^{\ast 2} & r^2 \zeta^{4} & r^2\\
    1 & r \zeta^3 & r \zeta^{\ast 3} & r^2\zeta^{6} & r^2    \end{array}
    \right|.  \,\,\,\,  ,  \zeta= e^{2\pi i/4}
    \label{D4star}
\end{equation}
The energy Eq.(\ref{EE}) of the 4-star is 
\begin{eqnarray}
{\cal E}_{4-star} (r) = - {\rm ln} r^{12} - {\rm ln}|\Delta_{4-star}(\zeta)|^2 + 4r^2, \,\,\,\,   \label{E4star} \\
\Delta_{4}(\eta)= \left| \begin{array}{ccccc} 1 & 0 & 0 & 0 & 0 \\
    1 & 1  &  1   &  1  & 1 \\
    1 &  \zeta &  \zeta^{\ast}  &  \zeta^2  & 1 \\
    1 &  \zeta^2 &  \zeta^{\ast 2} & \zeta^{4} & 1\\
    1 &  \zeta^3 &  \zeta^{\ast 3} & \zeta^{6} & 1    \end{array} \right| , 
 \,\,\,\, \zeta= e^{2\pi i /4}
\end{eqnarray}
The minimum of Eq.(\ref{E4star}) is at $r_{o}^2 = 3/2$. \\

In contrast, the ground state of a 5-cluster of spinless fermion in the fast rotating regime is the Verdermont determinant
$V({\bf z}) = || 1,z,z^2, z^3, z^4 ||$, 
\begin{equation}
    D_{5}^{QH}(z_1, .., z_5) = \left| \begin{array}{ccccc} 1 & z_1& z^{2}_{1} & z^{3}_{1} & z^{4}_{1}  \\
    1 & z_2& z^{2}_{2} & z^{3}_{2} & z^{4}_{2}\\
    1 & z_3& z^{2}_{3} & z^{3}_{3} & z^{4}_{3} \\
    1 & z_4& z^{2}_{4} & z^{3}_{4} & z^{4}_{4}\\
    1 & z_5& z^{2}_{5} & z^{3}_{5} & z^{4}_{5}
    \end{array}
    \right|.
    \label{V1}
\end{equation}
The energies in the 5-ring and 4-star families are 
\begin{eqnarray}
{\cal E}_{\rm 5-ring}^{QH} (r) = - {\rm ln} r^{20} - {\rm ln}|\Delta^{QH}_{\rm 5-ring}(\omega)|^2 + 5r^2, \,\,\,\,   \label{E5starQH} \\
\Delta^{QH}_{\rm 5-ring}(\eta)= \left| \begin{array}{ccccc} 1 & 1 & 1 & 1 & 1 \\
    1 & \omega  &  \omega^2   &  \omega^3  & \omega^4 \\
    1 & \omega^2  &  \omega^4   &  \omega^6  & \omega^8 \\
    1 & \omega^3  &  \omega^6   &  \omega^9  & \omega^{12}\\
    1 & \omega^4  &  \omega^8   &  \omega^{12}  & \omega^{16}   \end{array} \right| , 
 \,\,\,\, \omega= e^{2\pi i /5}; 
\end{eqnarray}

\begin{eqnarray}
{\cal E}_{\rm 4-star}^{QH} (r) = - {\rm ln} r^{20} - {\rm ln}|\Delta^{QH}_{\rm 4-star}(\zeta)|^2 + 4r^2, \,\,\,\,   \label{E4starQH} \\
\Delta^{QH}_{\rm 4-star}(\eta)= \left| \begin{array}{ccccc} 1 & 0 & 0 & 0 & 0 \\
    1 & 1  &  1   &  1  & 1 \\
    1 &  \zeta &  \zeta^{2}  &  \zeta^3  & \zeta^4 \\
     1 &  \zeta^2 &  \zeta^{4}  &  \zeta^6  & \zeta^8\\
    1 &  \zeta^3 &  \zeta^{6}  &  \zeta^9  & \zeta^{12}   \end{array} \right| , 
 \,\,\,\, \zeta= e^{2\pi i /4}.
\end{eqnarray}
The optimal radii in the 5-ring and 4-star family are $r^{2}_{o}= 2$ and  $r^{2}_{o}= 5/2$ respectively.  After evaluating the determinants $\Delta^{QH}_{\rm 5-ring}(\omega)$ and $\Delta^{QH}_{\rm 4-star}(\zeta)$, one finds that the minimum of the 5-ring has lower energy than that of 4-star.  In this way, we have constructed the Table in Figure 1 in the main text. \\

\noindent {\bf II.  Derivation of the principle curvature and principle vectors  of 3-clusters: }\\

\noindent {\bf (IIA) Slow rotation limit  ($\Omega=0^{+}$):} 
The ground state wavefunction is 
\begin{equation}
\Psi({\bf r}_{1}, {\bf r}_{2}, {\bf r}_{3}) = D({\bf r}_{1}, {\bf r}_{2}, {\bf r}_{3}) e^{-\sum_{i=1}^{3}|z_{i}|^2/2}, \,\,\,\, 
D({\bf r}_{1}, {\bf r}_{2}, {\bf r}_{3})  = \left( \begin{array}{ccc} 1 & z_{1} & z^{\ast}_{1} \\ 1 & z_{2} & z^{\ast}_{2}  \\
1 & z_{3} & z^{\ast}_{3} \end{array} \right). 
\label{Det-3}  \end{equation} 
Calculating optimal configuration within the 3-ring family as shown in the previous section, we find 
\begin{equation}
{\bf z}_{o} = r_{o}(1, \omega, \omega^2), \,\,\,\, 
r^{2}_{o}=  2/3, \,\,\,\,, \omega= e^{2\pi i/3}.    \label{ro}
\end{equation}
We, however, need to test the stability of this configuration in the full configuration space.  To do that, we look at the energy change near this configuration, 
\begin{equation}
{\cal E}({\bf r}_{o} + \delta {\bf r}) = {\cal E}({\bf r}_{o} ) + {\cal E}^{(1)}({\bf r}_{o}) +  {\cal E}^{(2)}({\bf r}_{o}) + \ldots
\end{equation}
where $ {\cal E}^{(1)}({\bf r}_{o})$ and ${\cal E}^{(2)}({\bf r}_{o}) $ are the first and second order changes in $\delta {\bf r}$. Explicitly, we have 
\begin{equation}
\Delta^{(1)}{\cal E} =  \delta z_{i} \frac{\partial {\cal E}}{\partial z_{i}} 
+  \delta z_{i}^{\ast}  \frac{\partial {\cal E}}{\partial z_{i}^{\ast}} 
\end{equation}
\begin{equation}
\frac{\partial {\cal E}}{\partial z_{i}}=  z^{\ast}_{i} - \left( \frac{\partial_{i} D}{D} + \frac{\partial_{i}D^{\ast}}{D^{\ast}} \right), 
\,\,\,\,  
\frac{\partial {\cal E}}{\partial z_{i}^{\ast}} =
  z_{i} - \left( \frac{\partial_{i}^{\ast} D}{D} + \frac{\partial_{i}^{\ast}D^{\ast}}{D^{\ast}} \right) 
\label{D1} \end{equation}
where $\partial_{i} \equiv \frac{\partial}{\partial z_{i}} =  (\partial_{x}- i \partial_{y})/2$,   
$\partial_{i}^{\ast} \equiv \frac{\partial}{\partial z_{i}^{\ast} } = (\partial_{x} + i \partial_{y})/2$.   
The second order change is 
\begin{equation}
{\cal E}^{(2)} =\frac{1}{2}{\cal Z}^{\dagger}_{i} {\cal W}_{ij} {\cal Z}_{j}, \,\,\,\,
 {\cal Z}_{i} = \left( \begin{array}{c} \delta z_{i} \\ \delta z_{i}^{\ast}\end{array} \right), \,\,\,\, 
 {\cal W}_{ij} =  
  \left( \begin{array}{cc} 
    M_{ij} & Q_{ij} \\ Q^{\ast}_{ij} & M^{\ast}_{ij}
    \end{array} \right)
\label{deltaE}    \end{equation}

\begin{equation}
 M_{ij} =\partial_{z_{i}^{\ast}}\partial_{z_{j}} {\cal E}({\bf z})= \delta_{ij} 
 - \left( \frac{\partial_{i}^{\ast}\partial_{j}^{} D }{D }  + \frac{\partial_{i}^{\ast}\partial_{j}^{} D^{\ast} }{D^{\ast} }   \right)
 + \left( \frac{\partial_{i}^{\ast}D \partial_{j}^{} D }{D^2 } + 
 \frac{\partial_{i}^{\ast}D \partial_{j}^{} D^{\ast} }{D^{\ast 2} }  \right)
 \end{equation} 
\begin{equation}
 Q_{ij} =\partial_{z_{i}^{\ast}}\partial_{z_{j}^{\ast}} {\cal E}({\bf z})= 
  - \left( \frac{\partial_{i}^{\ast}\partial_{j}^{\ast} D }{D } + 
 \frac{\partial_{i}^{\ast}\partial_{j}^{\ast} D^{\ast} }{D^{\ast} }  \right)
 + \left( \frac{\partial_{i}^{\ast}D \partial_{j}^{\ast} D }{D^2 } + 
 \frac{\partial_{i}^{\ast}D \partial_{j}^{\ast} D^{\ast} }{D^{\ast 2} }  \right).
 \label{Q}  \end{equation}

\vspace{0.2in}

{\em IIA.1. Global stability:} 
For slow rotation,  at the optimal configuration ${\bf r}_{o}$ in Eq.(\ref{ro}), we have 
\begin{equation} 
D({\bf r}_{o}) = 3(\omega^{\ast} - \omega), \,\,\,\,
\partial_{z_1} D({\bf z}) =  (z^{\ast}_{2} - z^{\ast}_{3} )_{{\bf r}_{o}} = \omega^{\ast}- \omega, 
\end{equation}
\begin{equation} 
D({\bf r}_{o})^{\ast} = 3(\omega - \omega^{\ast}), \,\,\,\,
\partial_{z_1} D({\bf z})^{\ast} =  (z^{\ast}_{3} - z^{\ast}_{2} )_{{\bf r}_{o}} = \omega- \omega^{\ast}. 
\end{equation}
We then have $\frac{\partial_{1} D}{D} = \frac{\partial_{1} D^{\ast}}{D^{\ast}}=1/3$. With
$(z_{1})_{o}$ in Eq.(\ref{ro}), we have 
\begin{equation}
\left( \frac{\partial {\cal E}}{\partial z_{1}} \right)_{{\bf r}_{o}} = 
\frac{2}{3} -(\frac{1}{3}+\frac{1}{3}) = 0. 
\end{equation}
Since all derivatives $\partial_{i}{\cal E}$ at the configuration Eq.(\ref{ro}) are related by rotation, we have 
$(\partial_{i} {\cal E} )_{{\bf r}_{o}} = 0$ for $i=1,2,3$. Hence,  ${\bf r}_{o}$ Eq.(\ref{ro}) is also the optimal configuration in the full configuration space. \\

{\em IIA.2. The principle curvatures:}  From Eq.(\ref{deltaE}) to Eq.(\ref{Q}), we can evaluate the matrix ${\cal W}$ in Eq.(\ref{deltaE}) at the optimal configuration ${\bf z}_{o}$ (Eq.(\ref{ro})).  We find

\begin{equation}
{\cal W} = \left( \begin{array}{cccccc} 1&0&0&0&0&0 \\ 0&1&0&0&0&0 \\
0&0&1&0&0&0 \\  0&0&0&1&0&0  \\0&0&0&0&1&0 \\ 1&0&0&0&0&1 \end{array} \right)
+ \frac{2}{(3r_{o})^2} 
 \left( \begin{array}{cccccc} 1&\omega & \omega^2 &1&\omega & \omega^2 \\ 
 \omega^2 &1&\omega & \omega &  \omega^2 &1 \\
 \omega &\omega^2 & 1 &\omega^2 &1&\omega  \\
 1&\omega^2 & \omega &  1&\omega^2 & \omega \\
 \omega^2 & \omega & 1 & \omega & 1 &\omega^2 \\
 \omega & 1 &\omega^2 & \omega^2 & \omega &  1
 \end{array}  \right)
 \end{equation}
where $r_{o}^2=2/3$.  To described the eigenvectors of ${\cal W}$, we introduce the vector
${\bf v}^{(0)}= (1, 1, 1)$, ${\bf v}^{(1)}= (1, \omega, \omega^2)/\sqrt{3}$,  ${\bf v}^{(2)}= (1, \omega^2, \omega)/\sqrt{3}$.  
Ordering the eigenvalues $\lambda_{n}$ in increasing order, we have 
\begin{eqnarray}
\lambda_{1}=0,   &\,\,\,\,\,\,  &\chi^{(1)} = (i {\bf v}^{(1)} , -i {\bf v}^{(1)\ast}   )   \\
\lambda_{2}=1,  & \,\,\,\,\,\,  &\chi^{(2)} = ({\bf v}^{(0)} ,  {\bf v}^{(0) \ast}  )   \\
\lambda_{3}=1,  & \,\,\,\,\,\, & \chi^{(3)} = ( i{\bf v}^{(0)} ,  -i {\bf v}^{(0) \ast}     )   \\
\lambda_{4}=2,  & \,\,\,\,\,\,  &\chi^{(4)} = ({\bf v}^{(1)} ,  {\bf v}^{(1) \ast}   )   \\
\lambda_{5}=2,   &\,\,\,\,\,\,  &\chi^{(5)} = ({\bf v}^{(2)} ,  {\bf v}^{(2) \ast}   )    \\
\lambda_{6}=2,   & \,\,\,\,\,\, & \chi^{(6)} = (i{\bf v}^{(2)} ,  -i{\bf v}^{(2) \ast}   )
\end{eqnarray}
Note that $i {\bf v}^{(1)} $ corresponds to a rotation of the optimal configuration (see Figure 2 in the main text).  It  has 0 eigenvalue because of the rotational invariance of the system. The vectors  ${\bf v}^{(0)}$ and $i {\bf v}^{(0)}$ corresponds to the translation along $x$ and $y$ direction. The mode ${\bf v}^{(1)}$ is the stretching mode. \\

For an $N$-ring, the eigenvectors $\chi^{(n)} = ({\bf v}^{(n)}, {\bf v}^{(n)\ast})$ are dictated by the rotational symmetry. They are of the form
\begin{equation} 
{\bf v}^{(p)}= (1, \zeta^{p}, \zeta^{2p}, ... , \zeta^{(N-1)p})/\sqrt{N}, \,\,\,\,\,\, p=0, 1, 2, ..., N-1.
\end{equation}
The vectors $(1,1, ... 1)$ and $i((1,1, ... 1)$ are the displacement of the entire pattern along the $x$ and $y$ directions. The vector $(1,\zeta,\zeta^2 ..., \zeta^{N-1})$ corresponds to stretching (or contracting) the optimal configuration along the symmetry directions. The vector $i(1,\zeta,\zeta^2 ..., \zeta^{N-1})$ corresponds to a rigid rotation, and is a zero energy mode. \\

\noindent {\bf (IIB) Fast rotation limit  ($\Omega=\omega^{-}_{h}$):}  In this limit the ground state has the determinant 
\begin{equation}
D({\bf r}_{1}, {\bf r}_{2}, {\bf r}_{3})  = \left( \begin{array}{ccc} 1 & z_{1} & z^{2} \\ 1 & z_{2} & z^{2}_{2}  \\
1 & z_{3} & z^{2}_{3} \end{array} \right). 
\end{equation}
The optimal configuration in the 3-ring family is $r_{o}^2=1$.  Repeating the  calculation for the slow rotation case, it is simple to show that 
$(\partial_{i}{\cal E})_{r_{o}}=0$, i.e. the configuration ${\bf z}_{o} = r_{o}(1, \omega, \omega^2)$ is a global minimum. 
The matrix ${\cal W}$ is given by 
\begin{equation}
{\cal W} = \left( \begin{array}{cccccc} 1&0&0&0&0&0 \\ 0&1&0&0&0&0 \\
0&0&1&0&0&0 \\  0&0&0&1&0&0  \\0&0&0&0&1&0 \\ 1&0&0&0&0&1 \end{array} \right)
+ \frac{1}{(3r_{o}^2)} 
 \left( \begin{array}{cccccc} 0& 0 & 0 &1&\omega & \omega^2 \\ 
 0 &0&0 & \omega &  \omega^2 &1 \\
 0 & 0 & 0 &\omega^2 &1&\omega  \\
 1&\omega^2 & \omega &  0 & 0 & 0 \\
 \omega^2 & \omega &  1 & 0 & 0 & 0 \\
 \omega & 1 &\omega^2 & 0  & 0 &  0
 \end{array}  \right).
 \end{equation}
 \vspace{0.2in}
 
 \vspace{0.2in}
 
 \noindent  {\bf III. Symmetry properties of the dimensionless energy {\cal E} and their applications}
 Because of rotation symmetry, and the mirror symmetry 
 $x\rightarrow x$, $y \rightarrow -y$ of the probability 
 $P([{\bf r}]) = |\Psi([{\bf r}])|^2$, we have 
 \begin{equation}
     {\cal E}({\bf z}) =  {\cal E}(\zeta{\bf z}) = 
     {\cal E}( {\bf z}^{\ast})
 \end{equation}
 where $\zeta$ is a phase factor that describes an arbitrary rotation, and $z\rightarrow z^{\ast}$ describes the mirror symmetry. In the main text, we have used the rotational symmetry to generate the $x$-aligned sample to bring close together the configurations in the original sample ${\cal S}$ that are rotational equivalent but have little overlap with each other.  One can also use the mirror symmetry to generate to quadruple the size of the sample by performing a mirror reflection along $x$ and along $y$ (i.e. changing $z^{(\alpha)}\rightarrow z^{(\alpha)\ast}$, and $z^{(\alpha)}\rightarrow  i z^{(\alpha)\ast}$) for all configuration $z^{(\alpha)}$ in the sample ${\cal S}$.  Combining the mirror symmetry and the rotational symmetry, the original sample with $M$ configurations will have $4NM$ configurations in the final $x$-aligned sample.

  \vspace{0.2in}
 
 \noindent {\bf IV. The optimal configurations of an N-cluster of spinless fermions as a function of $N$: } \\
 
 Minimizing the dimensionless energy ${\cal E}$ of an $N$-cluster numerically, we obtain its optimal configuration as a function of $N$ in both the slow and the fast rotation regime. The results are shown in the next two pages, labelled as ``harmonic oscillator" and ``quantum Hall" cases respectively. 

\newpage

\begin{figure}
\includepdf{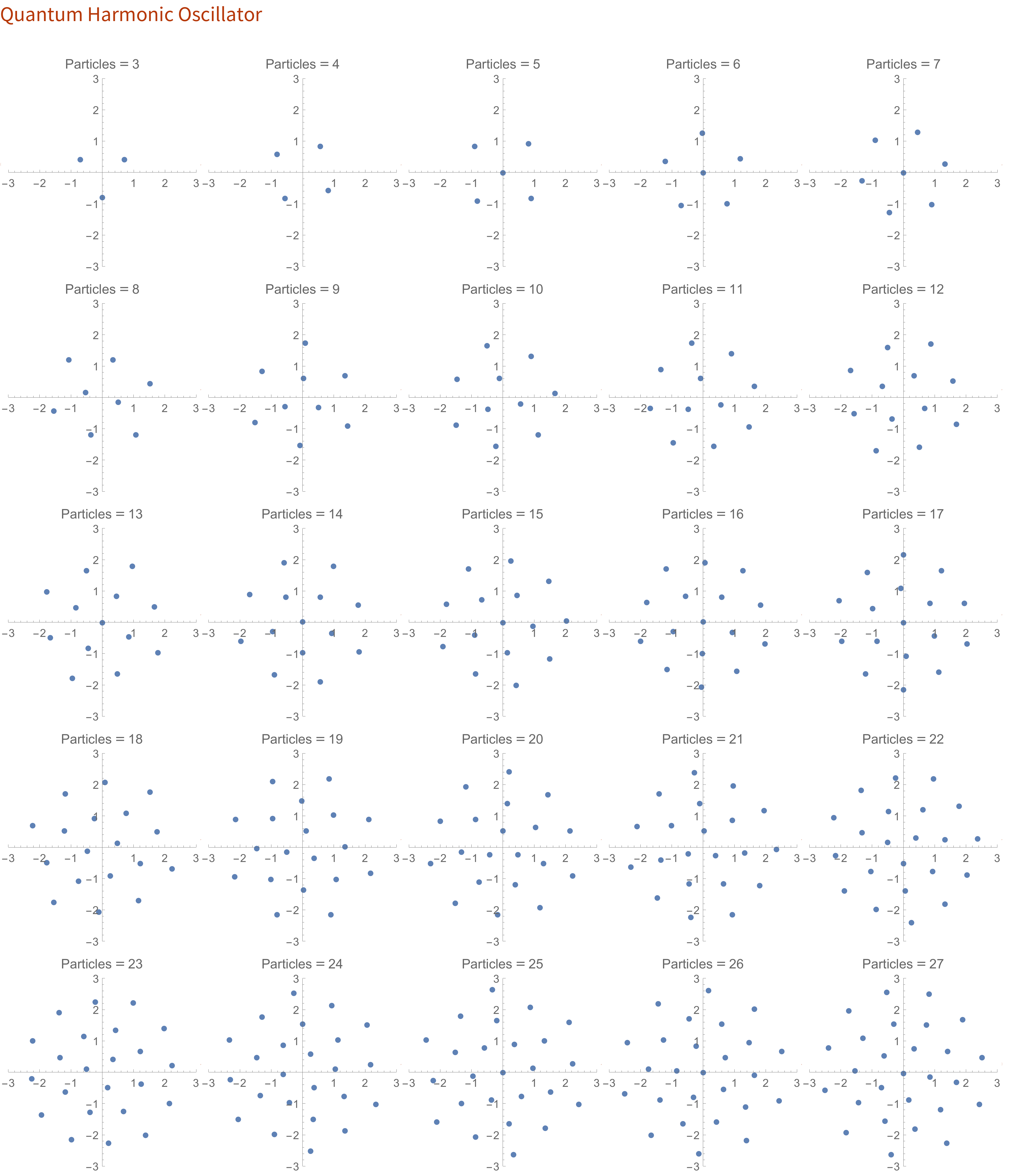}
\end{figure}

\clearpage
\newpage

\begin{figure}
\includepdf{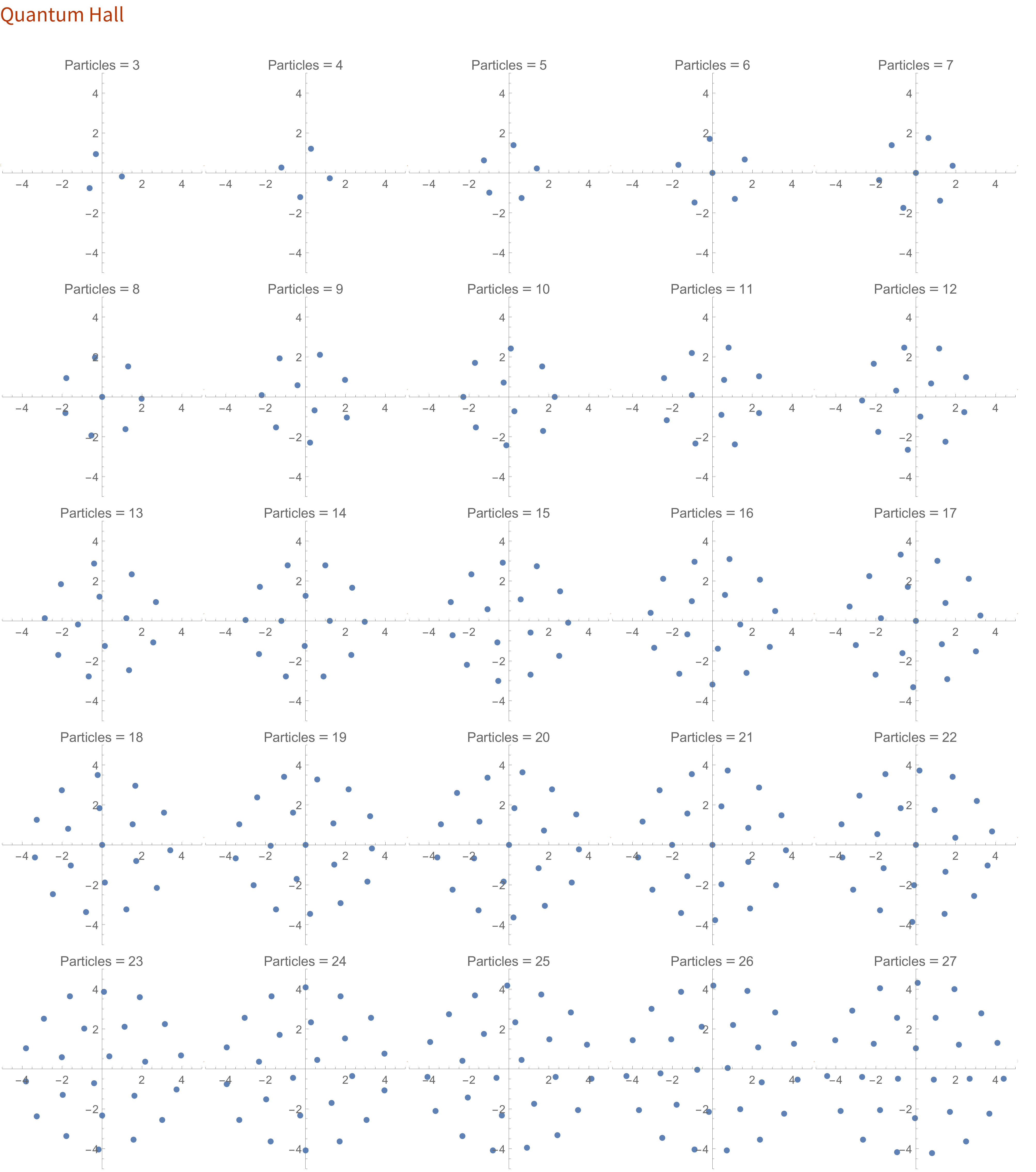}
\end{figure}

\end{document}